\documentclass[useAMS,natbib]{mn2e}
\usepackage{amssymb,amsmath}
\usepackage{graphicx}

\title[$R_{BLRs}$ of low luminosity AGN]
{the size of BLRs of low luminous Active Galactic Nuclei}

\author[Zhang et al.]
      {Xue-Guang, Zhang$^{1,2}$\thanks{xguang@astroscu.unam.mx},
       Dultzin-Hacyan, Deborah.$^1$,
       Ting-Gui, Wang$^2$ \\
      $^1$Instituto de Astronom\'ia, Universidad Nacional Aut\'onoma de
                M\'exico, Apdo Postal 70-264, M\'exico D. F. 04510, Mexico \\
      $^2$Center for Astrophysics, Department of astronomy and Applied
                Physics, University of Science and Technology of China, \\
                Hefei, Anhui, P.R.China}

\date{}

\def\LaTeX{L\kern-.36em\raise.3ex\hbox{a}\kern-.15em
   T\kern-.1667em\lower.7ex\hbox{E}\kern-.125emX}
\begin{document}
\label{firstpage}

\maketitle
\begin{abstract}
We study the size of BLRs of low luminosity Active Galactic
Nuclei, also called 'dwarf AGN', defined as having
($L_{H\alpha}\le10^{41}{\rm erg\cdot s^{-1}}$)). We more than
double the sample size analyzed previously (Wang \& Zhang 2003,
hereafter Paper I). In this study we first confirm our previous
result that the sizes of BLRs of low luminosity AGN are larger
than the ones expected from the empirical relation $R_{BLRs} -
L_{H\alpha}$ valid for 'normal' AGN: Seyfert 1s and quasars,
except for the objects with accretion rate
$\dot{m_{H\alpha}}>10^{-5.5}$. Second, we find a positive
correlation between the line width of the narrow emission line (as
tracer of velocity dipersion and thus bulge and black hole mass)
and the size of BLRs for both normal and low luminosity AGN. In
this paper we find a non-linear dependence of the BLRs sizes of low
luminosity AGN on BH masses. We also show that their sizes of BLRs are
more strongly dominated by the 'specific accretion rate'
$\dot{m_{H\alpha}}$ defined as $\dot{m_{H\alpha}} =
L_{H\alpha}/L_{Edd}$, than by the masses of their cetral black
holes. {  As an expected result, the distance of emission regions 
of low-ionization broad H$\alpha$ of NGC 4395 should be consistent with 
the value from the empirical relation of $R_{BLRs} - L_{H\alpha}$, 
according to the high accretion rate}

\end{abstract}

\begin{keywords}
Galaxies:Active -- Galaxies:nuclei -- Galaxies:low luminosity
\end{keywords}

\section{Introduction}

  Broad emission lines are one of the prominent properties of type 1 Active
Galactic Nuclei (AGN). However, due to the limit of the resolution
of present, and even forseeable astronomical instruments, we have
no way to resolve the geometric structure of broad emission line
regions (BLRs) by direct observations. Throughout this paper we
shall use the term 'size of BLRs' as equivalent of distance of
the BLRs to the central source. As the result of efforts of
hundreds of astronomers during more than ten years,  the sizes of
the BLRs have been estimated for more than 30 AGN 
by the Reverberation Mapping technique (Blandford \& Mckee
1982; Peterson 1993;  Kaspi et al. 2005). Using a sample of 35
objects, 15 PG quasars and 21 Seyfert 1 galaxies, a strong
correlation between the size of BLRs and the continuum
luminosity has been found (Kaspi et al. 2000; 2005). The
empirical relation $R_{BLRs} - L_{5100\AA}^{\alpha}$ indicates that
both the ionization continuum and the electron density of the BLRs
for normal AGN are not much different, and provides a convenient
way to estimate the central BH masses for AGN under the
assumption of Virialization (Greene \& Ho 2005b; Ovcharov et al.
2005; Wu et al. 2004; McLure \& Jarvis 2004; 2002; Marziani et
al. 2003), especially for objects with low and intermediate
redshifts and high S/N spectra (Sulentic et al. 2006). Due to the
difficulty to measure the stellar velocity dispersion, the
relation $M_{BH} - \sigma$ which can be extrapolated for
intermediate BH masses in the range of $10^4 - 10^6{\rm
M_{\odot}}$ (Barth et al. 2005; Greene \& Ho 2004), cannot be
applied to higher luminosity and higher redshift AGN (
Tremaine et al. 2002; Ferrarese \& Merritt 2001;
Gebhardt, Bender et al. 2000).

  In Paper I we have shown that the empirical relation
$R_{BLRs} - L_{H\alpha}^{\alpha}$ is not valid for a special kind
of AGN: low luminosity active galaxies with
$L_{H\alpha}<10^{41}{\rm erg\cdot s^{-1}}$ selected from Ho et al.
(1997a, 1997b), and called 'dwarf AGN' by them.  The deviation
of this empirical relation for low luminosity AGN may be explained
in terms of a lower ionization parameter and/or lower electron
density of BLRs in these type of AGN (although see discussion
paper I). Broad emission lines (often only ${H\alpha}$) can be
detected in the spectra of these type of galaxies indicating the
presence of activity in their nuclei, and there is also other
evidence for the presence of massive black holes in low luminosity
AGN (Nakai et al. 1993; Sarzi et al. 2001).  In the case of low
luminosity AGN with intermediate BH masses$M_{BH}\sim10^4-10^6{\rm
M_{\odot}}$, however,  several objects have been studied
individually in order to check if they follow the correlation
$M_{BH} - \sigma$: POX 52 (Barth et al. 2004), NGC 4395 (Kraemer
et al. 1999; Filippenko \& Ho 2003), NGC4051 (Shemmer et al.
2003) and various objects selected from SDSS (Greene \& Ho 2004).
The consistency between the BH masses estimated using the relation
$M_{BH} - \sigma$ and estimated under the assumption of
virialization, indicates that the size of BLRs for these AGN with
intermediate BH masses must obey the empirical relation $R_{BLRs} -
L_{5100\AA}$. Thus the question of whether all the sizes of BLRs
of low luminosity AGN obey the empirical relation is an
interesting one. In this paper, we use a larger sample of low
luminosity AGN to further investigate this question, in order to
find a more consistent explanation for the deviation of the
empirical relation for low luminosity AGN reported in Paper I. The
next section presents the sample. The results are shown in Section
III. Section IV gives the discussion and conclusion. In this
paper, the cosmological parameters $H_{0}=70{\rm km\cdot
s}^{-1}{\rm Mpc}^{-1}$, $\Omega_{\Lambda}=0.7$ and
$\Omega_{m}=0.3$ have been adopted.

\section{Sample}

  In a spectroscopic survey of more than four hundred nearby galaxies,
there are 34 AGN, of which the broad emission lines have been
detected with confidence, and 12 AGN with less certainty (Ho et
al. 1997a, 1997b). In paper I, we selected 22 low luminosity AGN
with broad emission H$\alpha$, 4 Seyfert galaxies and also 18
LINERs selected from the sample of Ho et al. (1997a, 1997b),
to study the size of BLRs. 
  
   As in paper I, we select low luminosity AGN according to the criterion
that $L_{H\alpha}\le10^{41}{\rm erg\cdot s^{-1}}$ (see also Ho et
al. 1997a; 1997b). We first select 42 low luminosity AGN from Ho
et al. (1997b), including the objects NGC4051 and NGC4395, for
which the Galactic and internal reddening have been corrected
using the values given by Ho et al. (1997a): The internal color
excess can be found in their Table 4, the Galactic color excess
can be found in their Table 11. Although there are 46 objects for
which broad H$\alpha$ lines have been detected, there are four
objects NGC1275, NGC4151, NGC3516 and NGC5548 with a luminosity of
H$\alpha$ larger than $10^{41}{\rm erg\cdot s^{-1}}$, thus, we
reject them for our sample. Furthermore, in paper I, there are 7
objects with double-peaked broad low-ionization emission lines
(dbp emitters), we excluded these objects from our new sample.
Finally, we also selected low luminosity AGN from the literature
according to the above defined criterion for $L_H\alpha$. We find
four additional objects which can be selected as candidate, POX 52
(Barth et al. 2004), SDSS J024912.86-081525.6, SDSS
J143450.62+033842.5 and SDSS J233837.10-002810.3 ({  Greene \& Ho
2004, 2005b}). In order to estimate the BH masses, we compile the
velocity dispersion of the bulge for each object. If there are no
data for the velocity dispersion of the bulge, we use the line
width of the strong narrow emission line [NII]$\lambda6583\AA$.
The strong correlation between the line width of the narrow
emission lines and the stellar velocity dispersion of the bulge
has been studied by several authors (e.g. Nelson \& Whittle 1996;
{  Greene \& Ho 2005a}). The sample objects are listed in Table 1.

 Finally, we select the 35 objects with accurate determinations for
the sizes of their BLRs estimated
from reverberation mapping technique, in order to compare the
difference for the two kinds of AGN, low luminosity and normal
AGN. The 35 normal AGN can be found in Kaspi et al. (2005) and
Peterson et al. (2004). We do not list them again.

\section{Results}
\subsection{The size of BLRs of dwarf AGN}
  As described in paper I, we estimate the size of BLRs of low luminosity
AGN using:
\begin{subequations}
\begin{align}
M_{BH} &= 10^{8.13\pm0.06}(\frac{\sigma}{200{\rm km\cdot s^{-1}}})^{4.02\pm0.32}
{\rm M_{\odot}}\\
R      &= q\times\frac{G\times M_{BH}}{FWHM_{H\alpha_{B}}^2}
\end{align}
\end{subequations}
where $FWHM_{H\alpha_{B}}$ is the line width of broad H$\alpha$,
and $q$ is a factor related to the orbits of the emission clouds.
It is also pointed out that $q=\frac{4}{3}$ is the most convenient
value discussed in the literature. In  more recent literature,
however, Onken et al. (2004) found that
$M_{BH}=5.5\times\frac{R\sigma_{line}^2}{G}$ and Peterson et al.
(2004) found the mean value of $FWHM/\sigma=2.03\pm0.59$. Thus,
here we take $q = \frac{1}{5.5/2.03^2}=0.75$ rather than $q =
\frac{4}{3}$.

  There are 38 low luminosity AGN, for which the stellar velocity
dispersions can be compiled from the recent literature. 
{\rm There is one object, NGC4395, of which the upper limit value 
of stellar velocity dispersion, $\sigma<30{\rm km\cdot s^{-1}}$, 
can be selected from Fillippenko \& Ho (2003).}
In order to estimate the BH masses for the other {\rm seven}, we first 
check the correlation between the stellar velocity dispersion and the line
width of narrow emission lines. {  We adopt the 
[NII]$\lambda6583\AA$ measurements of Ho et al. (1997a).} We
show this correlation in Fig 1. The Spearman Rank correlation
coefficient is 0.77 with $P_{null}\le8.8\times10^{-8}$. The
mean value of $FWHM/\sigma$ is about $2.13\pm0.12$. Here, we do
not consider the object NGC 1068, because of the special geometric
structure of its narrow emission line region (NLR), which has been
observed with HST. The NLR of NGC 1068 consists of a large set of
clouds (Macchetto et al. 1994; Capetti et al. 1995; Dietrich \&
Wagner 1998; Cecil et al. 2002) which leads to a larger than
mean value of line width for narrow emission lines. The strong
correlation seen in Fig 1 indicates that we can confidently
estimate the BH masses for the eight low luminosity AGN without
velocity dispersions using the line width of
[NII]$\lambda6583\AA$. Somewhat larger stellar velocity
dispersions than those derived from $\sigma_{[NII]}$ are perhaps
due to a larger distance of the [NII] emission line clouds to the
central source. Thus, the sizes of BLRs for these eight low
luminosity AGN can also be estimated. The sizes of the BLRs  are
listed in Table 1. {  Notice the case of the low luminosity AGN
NGC 4395 (because the size of BLRs has been measured by Reverberation Mapping 
technique by Peterson et al. (2005), thus we list the object individually) 
for which the BH mass estimated from the upper limit  
of stellar velocity dispersion is about $6.6\pm3.1\times10^4{\rm M_{\odot}}$ 
(we think the upper limit of stellar velocity from Keck high-resolution 
spectra is more reliable than the upper limit of line width of [NII]), 
which is about four to six times smaller than the BH mass 
$3.6\pm1.1\times10^5{\rm M_{\odot}}$ derived
recently under the assumption of Virialization and using
reverberation mapping by Peterson et al. (2005).}

  We can now compare the empirical relation $R_{BLRs} - L_{H\alpha}$ for
low luminosity AGN to that of normal AGN. As we can see from
Figure 2, the result from paper I is confirmed. The size of BLRs of
low luminosity AGN is larger than  the prediction of the empirical
relation $R_{BLRs} - L_{H\alpha}$. {  The mean value of $\xi =
\log(\frac{L_{H\alpha}/10^{44}{\rm erg\cdot
s^{-1}}}{(R_{BLRs}/100{\rm light-days})^2})$ is about
$-4.19\pm0.23$ for low luminosity AGN except NGC4395. 
The Spearman Rank correlation coefficient is 0.06
with $P_{null}\sim0.72$. For NGC 4395, the object mentioned
above, the size of the BLRs is about 0.04 light-days derived
directly from the variation of the CIV emission line, which is
much smaller than the upper limit value derived from the upper limit 
BH masses using broad H$\alpha$: 1.28 light-days. However, 
the emission regions of high-ionization CIV should be nearer to the 
central black hole 
than the regions of low-ionization H$\alpha$.} We do notice, 
however,  that there
are some low luminosity AGN including  four objects with
intermediate BH masses, for which the sizes of the BLRs do obey
the empirical relation $R_{BLRs} - L_{H\alpha}$. Thus, it is
important to further investigate which parameter dominates the
size of BLRs of both low luminosity and normal AGN.

\subsection{BH masses and accretion rates}

Due to the strong correlation between BH mass and the continuum
luminosity, there is also a strong correlation between BH mass and
the size of BLRs for normal AGN. We investigate if the same
holds for low luminosity AGN, which can be somehow expected given
that the sizes of the BLRs are also calculated from BH masses.

  The correlations between $M_{BH}$ and $L_{H\alpha}$ are shown in
Figure 3. {  The Spearman Rank correlation coefficient is -0.09 with
$P_{null}\ge51\%$ for low luminosity AGN except NGC4395}, and it 
is 0.87 with $P_{null}\sim1.15\times10^{-11}$ for normal AGN. Thus, we 
see that the correlations are different.  Next, we show the correlations
between $M_{BH}$ and $R_{BLRs}$ in Figure 4. {  The Spearman Rank
correlation coefficient is 0.94 with
$P_{null}\sim4.66\times10^{-22}$ for low luminosity AGN except NGC4395}, 
and 0.90 with $P_{null}\sim1.12\times10^{-13}$ for the 35 normal AGN. The
unweight linear relations for low luminosity AGN except NGC4395 
and normal AGN are:
\begin{subequations}
\begin{align}
&\log(R_{BLRs}) (low) = -4.31\pm1.14+0.78\pm0.15\log(M_{BH}) \\
&\log(R_{BLRs}) (normal) = -4.36\pm1.98+0.73\pm0.25\log(M_{BH})
\end{align}
\end{subequations}
A correlation between $M_{BH}$ and  $R_{BLRs}$ can be expected,
because of the fundamental formula $M_{BH} = q
\frac{R_{BLRs}FWHM_{H\alpha_{B}}^2}{G}$. However, the results above
show that there is no simple linear relation $R_{BLRs}\propto
M_{BH}^1$, instead $R_{BLRs}\propto M_{BH}^{\alpha}$ with
$\alpha<1$. Thus we can still say that there is a correlation
between BH masses and line width.

  We shall next define and calculate a parameter related to the
dimensionless accretion rate, $\dot{m} = L_{bol}/L_{Edd}$, where
$L_{bol}\sim9\times L_{5100\AA}$ is the bolometric luminosity for 
Normal AGN (Wandel et al. 1999; Kaspi et al. 2000) and $L_{Edd}$ is the
Eddington luminosity $L_{Edd} = 1.38\times10^{38}M_{BH}/{\rm
M_{\odot}erg\cdot s^{-1}}$. It is difficult to determine accurately
the continuum luminosity of the nucleus for low luminosity AGN,
because of the dominant component from stellar light and the lack of 
big blue bump. 

   Unfortunately, it was also shown by Ho (1999) that for low
luminosity AGN, the bolometric luminosity cannot be estimated as
$9\times L_{5100\AA}$, since for example the ratio of bolometric
luminosity to continuum luminosity at 5100$\AA$ for the low
luminosity AGN M81 (NGC 3031) is about 10, whereas the same ratio
for another low luminosity AGN, NGC 4261, is more than 500!.
Greene \& Ho (2005b) found a strong correlation between the
optical continuum luminosity and the luminosity of H$\alpha$
(including both broad and narrow components) using a sample of
normal AGN selected from SDSS, but we cannot be sure that the same
correlation applies to low luminosity AGN. In what follows, we use
the total (broad + narrow) luminosity of H$\alpha$ to define and
calculate another dimensionless accretion rate:
$\dot{m_{H\alpha}}$: $\dot{m_{H\alpha}} = L_{H\alpha}/L_{Edd}$.

  The correlation between the size of BLRs and the
dimensionless accretion rate $\dot{m_{H\alpha}}$ is shown in Figure 5.
{  The Spearman Rank correlation coefficient is -0.71 with
$P_{null}\le3.48\times10^{-8}$ for low luminosity AGN except NGC4395}. 
For the normal 35 AGN, we
select the 12 objects with measured stellar velocity dispersions to
calculate the dimensionless accretion rate $\dot{m_{H\alpha}} =
L_{H\alpha}/L_{Edd}$, and the luminosities of H$\alpha$ for the 12
objects are estimated from the continuum luminosity by the
correlation found by Greene \& Ho (2005b):
\begin{equation}
L_{H\alpha} = 5.25\pm0.02\times10^{42}(\frac{L_{5100\AA}}{{\rm 10^{44}
erg\cdot s^{-1}}})^{1.157\pm0.005} {\rm erg\cdot s^{-1}}
\end{equation}
The correlation coefficient is 0.76 with $P_{null}\le0.004$ for the
12 normal AGN. The unweighted linear relation for low luminosity
AGN is:
\begin{equation}
\log(R_{BLRs}) = -1.37\pm0.73 - 0.47\pm0.11\log(L_{H\alpha}/L_{Edd})
\end{equation}
Due to the small number of normal AGN, we cannot confirm
whether the different correlation between $R_{BLRs}$ and
$\dot{m_{H\alpha}}$ for low luminosity and normal AGN is really
true. However, we can confirm that the few low luminosity AGN
for which the sizes of the BLRs are consistent with the empirical
correlation $R_{BLRs} - L_{H\alpha}^{\alpha}$, have large values of
$\dot{m_{H\alpha}}$ than $\sim10^{-5.5}$. These low luminosity AGN
have been marked with names or shown as triangles in Figure 2.

\section{Discussion and Conclusions}

  BH masses estimated from stellar velocity dispersion are
reliable for low luminosity AGN. If the size of the BLRs of low
luminosity AGN with BH masses larger than $10^6{\rm M_{\odot}}$ would
obey the correlation $R_{BLRs} - L_{con}^{\alpha}$, their
lower continuum luminosity would lead to at least one order of magnitude
larger width of broad lines than what is observed in our sample. Such
broader emission lines cannot be detected in high resolution spectra
of HST. 
{  With respect to the issue of the
detectability limit of broad emission lines for ground based
observations,  we wish to point out as an example that Barth et
al. (2001) observed the BLRs for one low luminosity LINER: NGC
4579, with the STIS on HST, and detected broad H$\alpha$ with the
same width, $FWHM = 2300{\rm km\cdot s^{-1}}$, as the value
detected by Ho et al. (1997b). In the following section, we will
see that the line widths of broad H$\alpha$ in our sample selected
from Ho et al. (1997b) are not much less than 2300${\rm km\cdot
s^{-1}}$, thus we think these line widths are not limited by the
Palomar spectra detection capability. 
}
We cannot rule out however, the possibility that there may
be two distinct BLRs for at least some low luminosity AGN: The one
observed in the reported objects, and a very broad one near to the
black hole, possibly emitting double peaked lines, which is
presently undetectable.

   In this paper we have investigated the role of luminosity in the
structure of the BLR.  In Paper I it was found that low-luminosity
AGN  have abnormally large BLRs for a given luminosity, compared
to more luminous 'normal' AGN.  The current paper expands the
sample of low-luminosity AGN to include not only massive BHs in
low accretion states, but also intermediate-mass black holes
radiating close to their Eddington limits.  The latter, with
possibly one exception, despite their comparably low luminosities,
appear to have 'normal-sized' BLRs. The dependence of BLRs
structure on luminosity and Eddington ratio may have important
implications for BH mass measurements and AGN physics in general.
{  According to the high accretion rate of $\dot{m_{H\alpha}}$ of 
NGC4395, we can expect that the size of BLRs, particularly the emission 
regions of low-ionization braod H$\alpha$ should be consistent 
with the value from the empirical relation of $R_{BLRs} - L_{H\alpha}$.
}

\section*{Acknowledgements}
ZXG gratefully acknowleges the postdoctoral scholarships offered by la
Universidad Nacional Autonoma de Mexico (UNAM). D. D-H acknowledges
support from grant IN100703 from DGAPA, UNAM. This research has made use
of the NASA/IPAC Extragalactic Database (NED) which is operated by the Jet
Propulsion Laboratory, California Institute of Technology, under contract
with the National Aeronautics and Space Administration.This research has
also made use of the HyperLeda.

\begin{table*}
\centering
\begin{minipage}{175mm}
\caption{Data of Sample}
\begin{tabular}{llcccccl}
\hline
Name & Dis & $\log(H\alpha)$ & $\sigma$ & FWHM(H$\alpha$) & FWHM([NII]) &
$\log(R_{BLR})$ & Ref. \\
& Mpc & ${\rm erg\cdot s^{-1}}$ & ${\rm km\cdot s^{-1}}$ & ${\rm km\cdot s^{-1}}$ & ${\rm km\cdot s^{-1}}$ & light-day & \\
\hline
NGC315  &  65.8 & 39.87  &  310$\pm$3  &  2000  &  402  &  2.87 &  T98\\
NGC1052 &  17.8 & 40.25  &  215$\pm$7  &  1950  &  482  &  2.19 &  BHS02 \\
NGC1161 &  25.9 & 39.38  &  288        &  3000  &  518  &  2.45 &  M95\\
NGC2681 &  13.3 & 39.55  &  111$\pm$16 &  1550  &  265  &  1.31 & M95, OF91\\
NGC2787 &  13.0 & 38.76  &  202$\pm$5  &  2050  &  382  &  2.11 & BHS02\\
NGC2985 &  22.4 & 38.79  &  185$\pm$10 &  2050  &  263  &  1.95 &  S01\\
NGC3031 &   1.4 & 39.17  &  173$\pm$5  &  2650  &  210  &  1.61 &  Ve01\\
NGC3226 &  23.4 & 39.09  &  203$\pm$4  &  2000  &  465  &  2.14 &  PS02\\
NGC3642 &  27.5 & 39.46  &  109$\pm$11 &  1250  &  167  &  1.46 &  TDT\\
NGC3718 &  17.0 & 37.45  &  178$\pm$19 &  2350  &  371  &  1.77 & HS98 \\
NGC3998 &  21.6 & 40.57  &  297$\pm$5  &  2150  &  401  &  2.74 &  F97\\
NGC4036 &  24.6 & 39.17  &  166$\pm$5  &  1850  &  359  &  1.85 &  F97\\
NGC4143 &  17.0 & 39.13  &  211$\pm$6  &  2100  &  307  &  2.16 &  SP02\\
NGC4203 &   9.7 & 38.59  &  167$\pm$3  &  1500  &  346  &  2.05 &  BHS02\\
NGC4258 &   6.8 & 38.88  &   134       &  1700  &  296  &  1.29 &  p01\\
NGC4278 &   9.7 & 38.98  &  251$\pm$13 &  1950  &  479  &  2.53 & BHS02\\
NGC4450 &  16.8 & 38.83  &  80         &  2300  &  301  &  0.39 &  Z02\\
NGC4452 &  2.4  & 38.15  &  268$\pm$12 &  3700  &       &  2.09 &  SP97\\
NGC4579 &  16.8 & 39.45  &  165$\pm$4  &  2300  &  376  &  1.65 & BHS02\\
NGC4636 &  17.0 & 38.36  &  208$\pm$8  &  2450  &  333  &  2.01 &  BA02\\
NGC5077 &  40.6 & 39.23  &  275$\pm$39 &  2300  &  491  &  2.55 &  T98\\
NGC7479 &  32.4 & 39.97  &  109$\pm$11 &  2250  &  394  &  0.95 &  M95,TDT\\
NGC1068 &  14.4 & 40.38  &  254$\pm$10 &  3210  &  931  &  2.12 & GM97\\
NGC2639 &  42.6 & 40.07  &  193$\pm$13 &  3100  &  401  &  1.67 &  S83\\
NGC3227 &  20.6 & 40.98  &  128$\pm$13 &  3168  &  471  &  0.93 &  NW95\\
NGC3982 &  17.0 & 39.15  &  73$\pm$4   &  2150  &  227  &  0.29 &  BHS02\\
NGC4051 &  17.0 & 40.25  &  88$\pm$13  &  1072  &  228  &  1.22 &  NW95\\
NGC4138 &  17.0 & 38.65  &  100$\pm$10 &  1650  &  211  &  1.07 & JB96\\
NGC4168 &  16.8 & 38.32  &  187$\pm$8  &  2850  &  305  &  1.68 &  BA02\\
NGC4388 &  16.8 & 39.27  &  119$\pm$12 &  3900  &  280  &  0.63 &  TDT\\
NGC4565 &   9.7 & 38.42  &  144$\pm$7  &  1750  &  181  &  1.65 &  DB93\\
NGC4639 &  16.8 & 39.73  &  96$\pm$4   &  3600  &  180  &  0.32 &  BHS02\\
NGC5033 &  18.7 & 40.29  &  116$\pm$11 &  2850  &  261  &  0.85 &  BHS02\\
NGC5273 &  21.3 & 40.39  &  71$\pm$4   &  3350  &  147  &  -0.14& BHS02\\
NGC266  &  62.4 & 39.59  &             &  1350  &  326  &  1.99  &  \\
NGC841  &  59.5 & 39.76  &             &  1350  &  301  &  1.85  &  \\
NGC3884 &  91.6 & 40.25  &             &  2100  &  535  &  2.47  &  \\
NGC4395 &   3.6 & 38.08  &   30u       &  442   &  123u &  0.11u  &  FH03\\
NGC4438 &  16.8 & 39.47  &             &  2050  &  354  &  1.77  &  \\
NGC4750 &  26.1 & 39.12  &             &  2200  &  452  &  2.13  &  \\
NGC4772 &  16.3 & 38.75  &             &  2400  &  549  &  2.39  &  \\
NGC5005 &  21.3 & 40.05  &             &  1650  &  672  &  3.08  &  \\
POX52   &   93  & 40.06  &  36$\pm$5   &  765   &   87  &  -0.05 &  BARO4\\
GH02    & 129.2 & 40.29  &  59$\pm$8   &  690   &       &  0.91  &  BG05\\
GH14    & 122.5 & 40.29  &  59$\pm$7   &  770   &       &  0.81  &  BG05\\
GH19    & 158.4 & 40.06  &  53$\pm$4   &  1890  &       &  -0.16 &  BG05\\
\hline
\end{tabular}
\\
Notes:\\
The former 22 dwarf AGN are the objects in paper I, because, we
have used new value of stellar velocity dispersion from the more
recent
literature for some objects, thus, we also list them in the table. \\
{  For NGC4395, the stellar velocity dispersion is the upper limit 
value selected from Filippenko et al. (2003). The upper limit 
value of [NII] is selected from Ho et al. (1997a). \\}
{  References for Stellar velocity dispersion}:\\
BA02: Bernardi et al. 2002, BAR04: Barth et al. 2004, BHS02: Barth et al.
2002, BG05: Barth, Greene \& Ho 2005, DB93: DeSouza et al. 1993,
F97: Fisher 1997, FH03: Filippenko \& Ho 2003, GM97: Garcia-Lorenzo et al. 1997,
HS98: Heraudean \& Simien 1998, JB96: Jore et al. 1996,
M95: Mcelory 1995, NW95: Nelson \& Whittle 1995, OF91: Ore et al. 1991,
P01: Prugniel et al. 2001, PS02: Proctor \& Sansom 2002,
S83: Schechter 1983, S01: Sarzi et al. 2001, SP97: Simien \& Prugniel
1997,
SP02: Simien \& Prugniel 2002, T98: Trager et al. 1998,
TDT: Terlevich et al. 1990, ve01: Vega Beltran et al. 2001,
z02: de Zeeuw et al. 2002
\end{minipage}
\end{table*}

\begin{table*}
\centering
\begin{minipage}{115mm}
\caption{Data of the 12 objects}
\begin{tabular}{llllll}
\hline
Name & redshift & $\sigma$ & $L_{5100\AA}$ & $\log(M_{BH})$  & ref \\
    &          & ${\rm km\cdot s^{-1}}$ & ${\rm 10^{44}erg\cdot s^{-1}}$ & ${\rm M_{\odot}}$
& \\\hline
Mrk590  & 0.02638 & 169$\pm$28 &  0.653$\pm$0.069   & 7.84$\pm$0.36 & nw95 \\
3C120   & 0.03301 & 193$\pm$40 &  1.39$\pm$0.25     & 8.07$\pm$0.48 & sh90\\
Akn120  & 0.03230 & 177$\pm$20 &  1.69$\pm$0.12     & 7.92$\pm$0.26 & oo99\\
NGC3227 & 0.00386 & 128$\pm$13 &  0.0242$\pm$0.0028 & 7.35$\pm$0.46 & nw95 \\
NGC3516 & 0.00884 & 164$\pm$35 &  0.077$\pm$0.023   & 7.78$\pm$0.45 &  am97\\
NGC3783 & 0.00973 & 152$\pm$20 &  0.178$\pm$0.015   & 7.65$\pm$0.27 &  oo95\\
NGC4051 & 0.00234 & 88$\pm$13  &  0.0087$\pm$0.0004 & 6.69$\pm$0.21 &  nw95\\
NGC4151 & 0.00332 & 119$\pm$26 &  0.0815$\pm$0.0051 & 7.22$\pm$0.41 &  nw95\\
NGC4593 & 0.00900 & 209$\pm$20 &  0.122$\pm$0.039   & 8.21$\pm$0.25 & oo99 \\
IC4329A & 0.01605 & 225$\pm$20 &  0.208$\pm$0.026   & 8.33$\pm$0.24 &  oo99\\
NGC5548 & 0.01717 & 201$\pm$12 &  0.334$\pm$0.041   & 8.14$\pm$0.17 &  nw95\\
3C390.3 & 0.05610 & 273$\pm$16 &  0.87$\pm$0.34     & 8.67$\pm$0.22 &  le06\\
\hline
\end{tabular}
\\
Notes:\\
{  Column V is the BH masses estimated by the equation 1a.}\\
{  References for stellar velocity dispersion:\\} 
nw95: Nelson \& Whittle 1995; sh90: Smith, Heckman \& Illingworth 1990;
oo99: Oliva et al. 1999; am97: Arribas et al. 1997; le06: Lewis \&
Eraclous 2006
\end{minipage}
\end{table*}

\begin{figure}
\includegraphics[width=84mm]{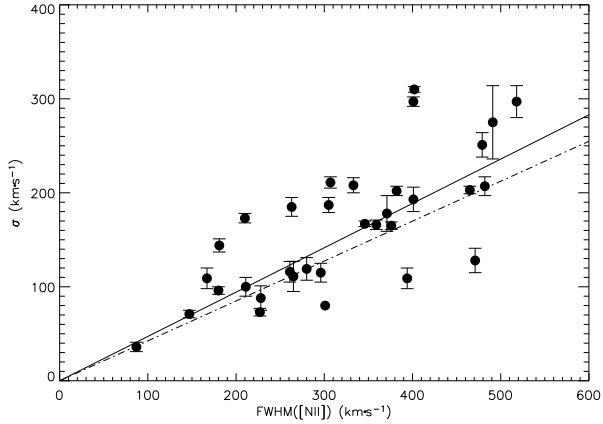}
\caption{The correlation between the stellar velocity dispersion $\sigma$
and the line width of [NII]$\lambda6583\AA$. The soild line represents
the relation $FWHM=2.12\sigma$. The dashed line represents the relation
$FWHM=2.35\sigma$}
\end{figure}

\begin{figure}
\includegraphics[width=84mm]{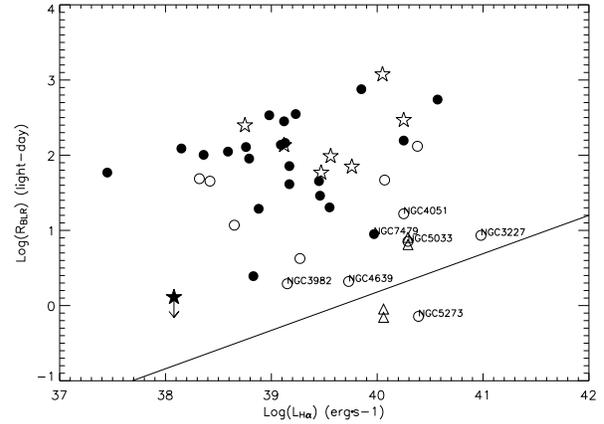}
\caption{The correlation between $R_{BLRs}$ and $L_{H\alpha}$. The
solid line represents the relation $\log(R_{BLRs}) =
0.51\log(L_{H\alpha}/10^{44}{\rm erg\cdot s^{-1}})+2.22$ found in paper I.
The solid circles are the objects in paper I. The open circles are
the ones with stellar velocity dispersions, but not included in
paper I. The five-point stars are  objects without stellar velocity
dispersions, for which the BH masses are estimated using the line
width of [NII]$\lambda6583\AA$. The solid five-point star represents
NGC 4395. The triangles are the four objects with intermediate BH
masses. The objects with larger values of $L_{H\alpha}/L_{Edd}$ are
marked with names or shown as triangle symbols.}
\end{figure}

\begin{figure}
\includegraphics[width=84mm]{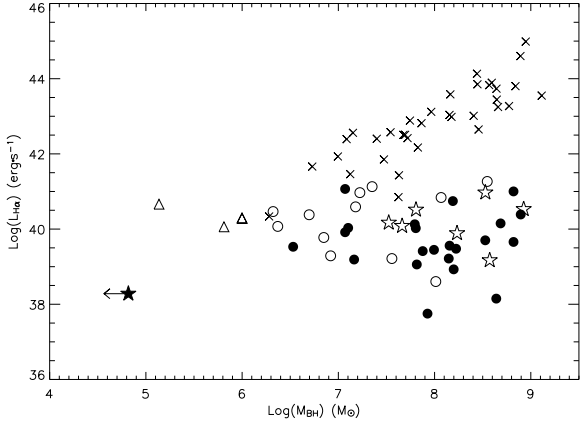}
\caption{The correlation between $M_{BH}$ and $L_{H\alpha}$. The
solid circles are the objects in paper I. The open circles are the
ones with stellar velocity dispersions, but not included in paper I.
The five-point stars are  objects without stellar velocity
dispersions, for which the BH masses are estimated using the line
width of [NII]$\lambda6583\AA$. The solid five-point star represent
NGC 4395. The triangles are the four objects with intermediate BH
masses. The crosses represent  normal AGN.}
\end{figure}

\begin{figure}
\includegraphics[width=84mm]{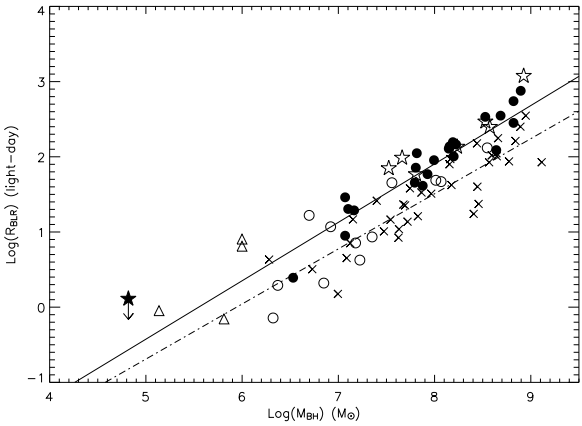}
\caption{The correlation between $M_{BH}$ and $R_{BLRs}$. The symbols
are the same as in Figure 3.}
\end{figure}

\begin{figure}
\includegraphics[width=84mm]{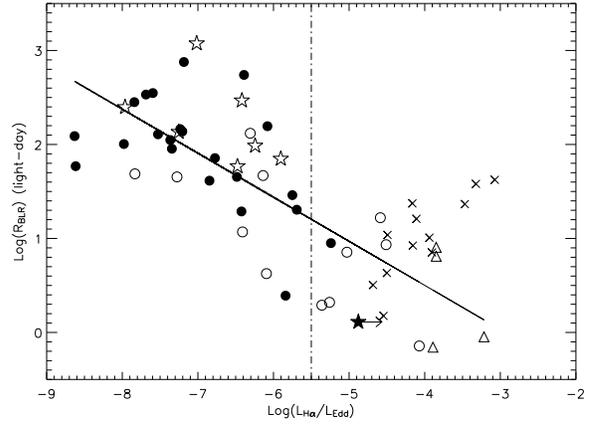}
\caption{The correlation between $R_{BLRs}$ and the value of
$L_{H\alpha}/L_{Edd}$. The symbols are the same as in Figure 3.}
\end{figure}


\begin{thebibliography}{}
\bibitem[]{am97}
Arribas S., Mediavilla E., Garcia-Lorenzo B., Del Burgo C., 1997, ApJ,
490, 227
\bibitem[Barth et al., 2001]{bh01}
Barth A. J., Ho L. C., Filippenko A. V. \& Rix H. W., 2001, ApJ, 546, 205
\bibitem[Barth et al., 2002]{bh02}
Barth A. J., Ho L. C. \& Sargent W. L. W., 2002, AJ, 124, 2607
\bibitem[Barth et al., 2004]{bar04}
Barth A. J., Ho L. C., Rutledge R. E. \& Sargent W. L. W., 2004, ApJ,
607, 90
\bibitem[Barth et al., 2005]{ba05}
Barth A. J., Greene J. E. \& Ho L. C., 2005, ApJ, 619, L151
\bibitem[Bernardi et al., 2002]{ba02}
Bernardi M., Alonso M. V., da Costa L. N., Willmer C. N. A., Wegner G.,
Pellegrini P. S., et al., 2002, AJ, 123, 2990
\bibitem[Blandford \& Mckee, 1982]{bm82}
Blandford R. D. \& Mckee C. F., 1982, ApJ, 255, 419
\bibitem[Capetti et al., 1995]{ca95}
Capetti A., Axon D. J., Macchetto F. D., Sparks W. B. \& Boksenberg A.,
1995, APJ, 446, 155
\bibitem[Cecil et al., 2002]{cd02}
Cecil G., Dopita M. A., Groves B. et al., 2002, APJ, 568, 627
\bibitem[DeSouza et al., 1993]{db93}
DeSouza R. E., Barbuy B. \& dos Anjos S., 1993, AJ, 105, 1737
\bibitem[Dietrich \& WAGNer, 1998]{dw98}
Dietrich M. \& Wagner S. J., 1998, A\&A, 338, 405
\bibitem[Ferrarese \& Merritt, 2001]{fm01}
Ferrarese L. \& Merritt D., 2001, MNRAS, 320, L30
\bibitem[Filippenko \& Ho, 2003]{fo03}
Filippenko A. V. \& Ho L. C., 2003, ApJ, 588, L13
\bibitem[Fisher, 1997]{F97}
Fisher D., 1997, AJ, 113, 950
\bibitem[Garcia-Lorenzo et al., 1997]{gm97}
Garcia-Lorenzo B., Mediavilla E., Arribas S. \& Del Burgo C., 1997, ApJ,
483, L99
\bibitem[Gebhardt, Bender et al., 2000]{geb00a}
Gebhardt K., Bender R., Bower G., et al., 2000, ApJ, 439, L13
\bibitem[Green \& Ho, 2004]{gh04}
Greene J. E. \& Ho L. C., 2004, ApJ, 610, 722
\bibitem[Green \& Ho, 2005a]{gh05a}
Greene J. E. \& Ho L. C., 2005a, ApJ, 627, 721
\bibitem[Green \& Ho, 2005b]{gh05b}
Greene J. E. \& Ho L. C., 2005b, ApJ, 630, 122
\bibitem[Heraudean \& Simien, 1998]{hs98}
Heraudean P. H. \& Simien F. 1998, A\&AS, 133, 317
\bibitem[Ho et al., 1997a]{ho97a}
Ho L. C., Filippenko A. \& Sargent W. L. W., 1997a, ApJS, 112, 315
\bibitem[Ho et al., 1997b]{ho97b}
Ho L. C., Filippenko A. \& Sargent W. L. W., 1997b, ApJS, 112, 391
\bibitem[Ho, 1999]{ho99}
Ho L. C., 1999, ApJ, 516, 672
\bibitem[Jore et al., 1996]{jb96}
Jore K.P., Broeils A.H. \& Haynes M.P., 1996, AJ, 112, 438
\bibitem[Kaspi et al., 2000]{kas00}
Kaspi S., Smith P. S., Netzer H., et al., 2000, ApJ, 533, 631
\bibitem[Kaspi et al., 2005]{kas05}
Kaspi S., Maoz D., Netzer H., Peterson B. M., et al., 2005, ApJ, 629, 61
\bibitem[Kraemer et al., 1999]{kr99}
Kraemer S. B., Ho L. C., Crenshaw D. M., et al., 1999, ApJ, 520, 564
\bibitem[Lewis \& Eracleous, 2006]{le06}
Lewis K. T. \& Eracleous M., 2006, astro-ph/0601398
\bibitem[Macchetto et al., 1994]{ma94}
Macchetto F. D., Capetti A., Sparks W. B., Axon D. J. \& Boksenberg A.,
1994, APJ, 435, L15
\bibitem[Marziani et al., 2003]{ma03}
Marziani P., Zamanov R. K., Sulentic J. W. \& Calvani M., 2003, MNRAS,
345, 1133
\bibitem[Mcelory, 1995]{m95}
Mcelory D. B., 1995, ApJS, 100, 105
\bibitem[McLure \& Jarvis, 2002]{mj02}
McLure R. J. \& Jarvis M. J., 2002, MNRAS, 337, 109
\bibitem[McLure \& Jarvis, 2004]{mj04}
McLure R. J. \& Jarvis M. J., 2004, MNRAS, 353, 45
\bibitem[Nakai et al., 1993]{na93}
Nakai N., Inoue M. \& Miyoshi M., 1993, Nature, 361, 45
\bibitem[Narayan et al., 1995]{na95}
Nelson C. H. \& Whittle M., 1995, ApJS, 99, 67
\bibitem[Nelson \& Whittle, 1996]{nw96}
Nelson C. H. \& Whittle M.,1996, APJ, 465, 96
\bibitem[]{oo99}
Oliva E., Origlia L., Maiolino R. \& Moorwood A. F. M., 1999, A\&A, 350,
9
\bibitem[Onken et al., 2004]{on04}
Onken C. A., Ferrarese L., Merritt D., Peterson B. M., et al., 2004,
ApJ, 615, 645
\bibitem[Ore et al., 1991]{of91}
Ore C. D., Faber S. M., Gonzalez J., Stoughton R. \& Burstein D., 1991,
ApJ, 366, 38
\bibitem[Ovcharov et al., 2005]{ov05}
Ovcharov E., Ivanov V. D., Nedialkov P., 2005, gbha.conf, 134O
\bibitem[Peterson, 1993]{pe93}
Peterson B. M., 1993, PASP, 105, 247
\bibitem[Peterson et al., 2004]{pe04}
Peterson B. M., Ferrarese L., Gilbert K. M., Kaspi S., et al., 2004,
ApJ, 613, 682
\bibitem[Peterson et al., 2005]{pe05}
Peterson B. M., Bentz M. C., Desroches L. B., et al., 2005, ApJ, 632, 799
\bibitem[Proctor \& Sansom, 2002]{ps02}
Proctor R. N. \&  Sansom A. E., 2002, MNRAS, 333, 517
\bibitem[Prugniel et al., 2001]{p01}
Prugniel P., Maubon G. \& Simien F., 2001, A\&A, 366, 68
\bibitem[Sarzi et al., 2001]{sa01}
Sarzi M., Rix H., Shields J. C., et al., 2001, ApJ, 550, 65
\bibitem[Schechter, 1983]{s83}
Schechter P. L. 1983, ApJS, 52, 425
\bibitem[Shemmer et al., 2003]{sh03}
Shemmer O., Uttley P., Netzer H. \& Mchaardy I. M., 2003, MNRAS, 343,
1341
\bibitem[Simien \& Prugniel, 1997]{sp97}
Simien F. \& Prugniel P., 1997, A\&AS, 126, 15
\bibitem[Simien \& Prugniel, 2002]{sp02}
Simien F. \& Prugniel P., 2002, A\&A, 384, 371
\bibitem[]{sh90}
Smith E. P., Heckman T. M. \& Illingworth G. D., 1990, ApJ, 356, 399
\bibitem[Sulentic et al., 2006]{su06}
Sulentic J. W., Repetto P., Stirpe G. M., Marziani P., Dultzin-Hacyan D.
\& Calvani M., 2006, A\&A, in press
\bibitem[Terlevich et al., 1990]{tdt90}
Terlevich E., Diaz A. I. \&  Terlevich R. 1990, MNRAS 242, 271
\bibitem[Trager et al., 1998]{t98}
Trager S. C., Worthey G., Faber S. M., et al., 1998, ApJS, 116, 1
\bibitem[Tremaine et al., 2002]{tr02}
Tremaine S., Gebhardt K., Bender R., Bower G., et al., 2002, ApJ, 574,
740
\bibitem[Vega et al., 2001]{ve01}
Vega Beltran J. C., Pizzella A., Corsini E. M., Funes J. G., Zeilinger
W. W., Beckman J. E. \& Bertola F., 2001, A\&A, 374, 394
\bibitem[Wandel, et al., 1999]{wan99}
Wandel A., Peterson B. M. \& Malkan M. A., 1999, ApJ, 526, 579
\bibitem[Wang \& Zhang, 2003]{wz03}
Wang T. G. \& Zhang X. G., 2003, MNRAS, 340, 793
\bibitem[Wu et al., 2004]{wu04}
Wu X. B., Wang R., Kong M. Z., et al., 2004, A\&A, 424, 793
\bibitem[de Zeeuw et al., 2002]{z02}
de Zeeuw P. T., et al., 2002, MNRAS, 329, 513
\end{thebibliography}
\end{document}